\documentclass[%
 reprint,
superscriptaddress,
 amsmath,amssymb,
 aps,
]{revtex4-2}
\usepackage{wrapfig}
\usepackage{graphicx}
\usepackage[table]{xcolor}
\usepackage[utf8]{inputenc} 
\DeclareUnicodeCharacter{2212}{-}
\usepackage{dcolumn}
\usepackage{bm, bbm, todonotes}
\newcommand{\be}{\begin{equation}}
\newcommand{\ee}{\end{equation}}

\usepackage{physics}
\usepackage{booktabs}
\usepackage{amssymb}
\usepackage{amsmath}
\usepackage{amsfonts}
\usepackage[colorlinks=true,linkcolor=blue,allcolors=blue]{hyperref}
\usepackage{graphicx}

\usepackage{algorithm}
\usepackage{algpseudocode}
\usepackage{optidef}
\usepackage{mathrsfs}
\usepackage{hyperref}

\def\R{\mathbb{R}}

\DeclareRobustCommand{\bigO}{%
  \text{\usefont{OMS}{cmsy}{m}{n}O}%
}

\begin{document}

\newcommand{\zz}{$\mathbb{Z}_2\ $}


\title{Trapped-ion quantum simulation of the Fermi-Hubbard model as a \\lattice gauge theory using hardware-aware native gates}

\author{Dhruv Srinivasan}
 \email{dhruv404@mit.edu}
\affiliation{Department of Mechanical Engineering, University of Maryland, College Park, MD 20742, USA.}
\affiliation{Department of Physics, University of Maryland, College Park, MD 20742, USA.}

\author{Alex Beyer}%
\affiliation{Department of Mechanical Engineering, University of Maryland, College Park, MD 20742, USA.}

\author{Daiwei Zhu}%
\affiliation{%
 IonQ, Inc., College Park, MD 20740, USA.
}%

\author{Pranav Srikanth}%
\affiliation{Department of Mechanical Engineering, University of Maryland, College Park, MD 20742, USA.}
\affiliation{Birla Institute of Technology and Science, Pilani, Rajasthan 333011, India}

\author{Spencer Churchill}%
\affiliation{%
 IonQ, Inc., College Park, MD 20740, USA.
}%

\author{Kushagra Mehta}%
\affiliation{Department of Physics, University of Maryland, College Park, MD 20742, USA.}

\author{Sashank Kaushik Sridhar}%
\affiliation{Department of Mechanical Engineering, University of Maryland, College Park, MD 20742, USA.}

\author{Kushal Chakrabarti}%
\affiliation{%
Tata Consultancy Services Research, Mumbai 400607, India.
}%

\author{David W. Steuerman}%
\affiliation{%
IonQ, Inc., College Park, MD 20740, USA.
}%
\affiliation{National Quantum Laboratory (QLab) at Maryland, College Park, MD 20742, USA.}

\author{Nikhil Chopra}%
 \email{nchopra@umd.edu}
\affiliation{Department of Mechanical Engineering, University of Maryland, College Park, MD 20742, USA.}

\author{Avik Dutt}%
 \email{avikdutt@umd.edu}
\affiliation{Department of Mechanical Engineering, University of Maryland, College Park, MD 20742, USA.}
\affiliation{Institute for Physical Science and Technology, University of Maryland, College Park, MD 20742, USA.}
\affiliation{National Quantum Laboratory (QLab) at Maryland, College Park, MD 20742, USA.}


\begin{abstract}

The Fermi-Hubbard model (FHM) is a simple yet rich model of strongly interacting electrons with complex dynamics and a variety of emerging quantum phases. These properties make it a compelling target for digital quantum simulation. While Trotterization-based quantum simulations have shown promise, implementations on current hardware are limited by noise, necessitating error mitigation techniques like circuit optimization and post-selection. A mapping of the FHM to a \zz LGT was recently proposed that restricts the dynamics to a subspace protected by additional symmetries, but a faithful representation of its dynamics on quantum hardware remains a challenge. In this work, we propose and demonstrate a suite of algorithm-hardware co-design strategies on a trapped-ion quantum computer, targeting two key aspects of NISQ-era quantum simulation: circuit compilation and error mitigation. To this end, we apply iteratively preconditioned gradient descent (IPG), which reduces the 2-qubit gate count of the FHM simulation by 14\%. When combined with von Neumann entropy (VNE) compression, this results in a total reduction of 36\%. We observe in QPU experiments that IPG alone, with error mitigation based on conserved symmetries (debiasing and sharpening techniques), enables a faithful observation of domain wall dynamics —even in the presence of noise that attenuates the magnitude of these overall dynamics. Using IPG in tandem with VNE compression in a noisy classical statevector simulator, we faithfully double the number of simulatable Trotter steps. Our work demonstrates the value of algorithm-hardware co-design to operate digital quantum simulators at the threshold of maximum circuit depths allowed by current hardware, and is broadly generalizable to strongly correlated systems in quantum chemistry and materials science.

\end{abstract}

\maketitle


\section{\label{sec:level1}Introduction}

The Fermi-Hubbard Model (FHM) is a simple yet powerful model that brings insight into materials' electronic and magnetic properties \cite{hubbard_electron_1997, esslinger_fermi-hubbard_2010}. It is promising in studies of quantum magnetism, antiferromagnetism, high-temperature superfluidity, d-wave pairing and superconductivity, and transitions between these different regimes of the rich phase diagram \cite{SCALAPINO1995329, hofstetter_high-temperature_2002, anderson_resonating_1987, anderson_model_1975}. 
Although analog quantum simulation based on ultracold-atom, trapped-ion and semiconductor-quantum-dot experiments have shown significant progress in quantum simulation of the FHM \cite{hebbe_madhusudhana_benchmarking_2021, hensgens_quantum_2017, davoudi_towards_2020, lebrat_observation_2024, hofrichter_direct_2016, taie_observation_2022}, such explorations are typically limited to regions defined by the experiment setup. In contrast, digital quantum simulation is universal in nature, but in practice challenging with today's noisy intermediate-scale quantum (NISQ) devices. There are many paths to address these challenges.

It is possible to encode the simulation into a symmetry protected subspace. A lattice gauge theory (LGT) approach for digital quantum simulation of the one-dimensional FHM has been recently proposed and analyzed \cite{khodaeva_quantum_2024}. Although requiring more qubits and deeper circuits than a direct encoding, the LGT approach allows for a set of local operators on the links between the lattice sites to be fixed to eigenvalues of $\pm 1$ depending on the sign of the interaction term in the Hubbard Model. It is typical for LGTs to introduce additional conservation laws that restrict time evolution to a small sector of the entire Hilbert space \cite{norbert_symmetry}. For example, in quantum link models for lattice quantum electrodynamics \cite{chandrasekharan_quantum_1997, martinez_real-time_2016, barbiero_fixed, shaw_quantum_2020, banerjee_atomic_2012, wiese_ultracold_2013, chanda_confinement_2020, surace_lattice_2020}, time evolution is naturally or artificially restricted to the sector that respects Gauss's law, and this has been used for post-selection error mitigation \cite{yang_observation_2020, schweizer_fixed, FRANK2020135484, yang_analog_2016}. In a similar spirit, the \zz LGT encoding of the FHM, in addition to global spin conservation, introduces a large number of local conserved charges \cite{khodaeva_quantum_2024}, which can enable efficient post-selection error mitigation. 

Circuit compilation is another effective technique. More generally, today’s NISQ era digital quantum computers, although large enough to conduct nontrivial gate operations, are limited by qubit coherence time, larger-than-ideal gate execution time, noise in operations, and differing qubit basis sets dependent on the qubit platform. Therefore, minimizing the depth of transpiled circuits is critical in modeling challenging systems such as the Fermi-Hubbard model.

Directly executing Trotterized FHM simulations encoded as a \zz LGT on current quantum processing units (QPUs) remains infeasible without first optimizing each Trotter step. A parametrized circuit template, known as an ansatz, can be used to optimize each Trotter step. However, classically parameterizing and optimizing a Trotter step ansatz is infeasible beyond a small number of qubits, due to the Hilbert space of a circuit increasing exponentially with the number of qubits. Such a problem is ubiquitous in variational algorithms \cite{schuld_effect_2021, verchere_fixed, d_quantum_2024, bravo-prieto_variational_2020}, including calculating ground state energies on a variational quantum eigensolver \cite{kandala_hardware-efficient_2017, grimsley_adaptive_2019, tilly_variational_2022, bentellis_fixed}. Even if a Trotter step can be classically optimized, ensuring that an ansatz possesses the expressibility needed to represent it requires attempting different numbers of single- and 2-qubit gates and their arrangements, and optimizing each gate angle. With a large number of parameters and no guarantee that a specific gate arrangement is sufficiently expressible, commonly used gradient-based optimizers struggle to converge to a solution (this is elaborated on in Sec. \ref{sec:opt_ansatz}). Additionally, transpiling Trotter steps made up of gates non-native to a QPU can significantly increase overall gate count, further rendering a circuit infeasible to execute. 

Here, we address these challenges by exploiting the symmetry of an objective unitary to allow for classically feasible circuit optimization. Specifically, we use the Trotter step of a Fermi-Hubbard Model via \zz LGT transformation as our target unitary and decompose this unitary as a combination of repeating, static three-qubit circuits. We then create a hardware-aware, parameterizable, three-qubit ansatz using the native gateset of the IonQ Aria QPU, and optimize this ansatz using a newly developed optimization technique –- that of iteratively pre-conditioned gradient descent (IPG)~\cite{kushal20} –- and compare the effectiveness of IPG to that of commonly used gradient-based optimizers. Finally, we run the circuits on the IonQ Aria QPU, leveraging debiasing and sharpening as error mitigation techniques combined with the conserved-symmetry-based error mitigation approach inherent to lattice gauge theories. 

This novel, systematic method of approaching circuit optimization on a Trotterized system results a 14\% reduction in the number of native two-qubit gates (Appendix, Sec. \ref{sec:ionq_QC}) per Trotter step, combined with an even larger reduction in the number of single qubit gates. 
A further reduction of 22\% is effected by using von Neumann entropy based arguments for reducing circuit depth while maintaining sufficient expressibility, resulting in a total 36\% reduction in 2-qubit gate count.
This, combined with debiasing and symmetry-based error mitigations, demonstrates a strategy to reliably run a large depth circuit of a  fermionic system on present day QPU hardware.

The structure of the paper is the following. In Sec. \ref{sec:encoding} we introduce and outline the quantum circuit of the Trotterized \zz LGT Fermi Hubbard Model. In Sec. \ref{sec:parameterized} we decompose the circuit into three-qubit sub-circuits and create a thirty-parameter, hardware-aware ansatz with adequate expressibility. In Sec. \ref{sec:ipg} we introduce iteratively pre-conditioned gradient descent (IPG) in the context of our specific quantum circuit optimization. In Sec. \ref{sec:opt_ansatz} we optimize the sub-circuit with IPG, and combine sub-circuits to create an overall optimized Trotter step. In Sec. \ref{sec:von_Neumann} we create a further optimized circuit via von Neumann entropy compression. In sec. \ref{sec:error-mitigation} we introduce the several error mitigation techniques we employed. In Sec. \ref{qpu-res} we run multiple trotter steps on the IonQ Aria QPU with Debiasing and Sharpening as active error mitigation strategies, but without VNE compression. In Sec. \ref{classical-res} we run multiple trotter steps optimized with VNE compression on a depolarizing noise statevector simulator. We conclude in Sec. \ref{sec:conclusions}.

\section{\label{sec:encoding}Encoding the FHM as a lattice gauge theory}
\subsection{\label{sec:level2}Fermi Hubbard Model}
The Fermi-Hubbard model (FHM) is the simplest Hamiltonian describing interacting electrons on a periodic lattice, which in the second quantized form reads \cite{stanisic_observing_2022, hubbard_electron_1997, esslinger_fermi-hubbard_2010, gebhard_mott_1997},
%
\begin{equation}
    H = - \sum_{\langle i, j \rangle \sigma} J_{ij, \sigma} \hat c_{i \sigma}^\dagger \hat c_{j \sigma}  + {\rm H.c.} + \frac{U}{2} \sum_{i} (\hat n_{i \uparrow} +  \hat n_{i \downarrow} - 1)^2 \label{eq:FH}
\end{equation}
where the first term denotes kinetic energy with nearest neighbor hopping strength $J_{ij, \sigma}$, and the second term denotes on-site interaction energy with strength $U$. $\hat{c}_{i \sigma}^{\dagger}$ and $\hat{c}_{i \sigma}$ are the creation and annihilation operators respectively for a spin-1/2 fermion with $\sigma \in \{\uparrow, \downarrow\}$, and ${\hat n_{i \sigma}} = \hat{c}_{i \sigma}^{\dagger}\hat{c}_{i \sigma}$ is the number operator on site $i$. H.c. denotes the Hermitian conjugate. Fermionic anticommutation is obeyed via $\{\hat{c}_{i \sigma}^{\dagger}\hat{c}_{i \sigma'} \} = \delta_{i,j}\delta_{\sigma, \sigma'}$. The form in Eq. \eqref{eq:FH} enforces a special unitary group (SU(2)) algebra due to the spin-1/2 nature of the fermions, although recent experiments have probed SU(N) generalizations of the FH model \cite{hofrichter_direct_2016, taie_observation_2022}.

The standard way to map the FHM on a universal gate-based quantum computer involves the choice of a specific spin representation combined with a Jordan-Wigner transformation to convert the fermion operators to qubits \cite{giamarchi2003quantum, khodaeva_quantum_2024, jordan_uber_1928}. Here we choose the so-called slave-spin representation and a 1D lattice for the subsequent Jordan-Wigner transformation since that preserves the nearest-neighbor (local) nature of the couplings in the Hamiltonian of Eq. \eqref{eq:FH} \cite{ruegg_z_2010, essler_one-dimensional_2005}. Note that higher-dimensional FHM lattices require long-range nonlocal couplings, which can be implemented in the trapped-ion QC architecture we discuss in Secs. \ref{sec:ionq_QC} since this QC hardware platform possesses all-to-all qubit connectivity without requiring SWAP gates. Additionally, this means that the geometry of a lattice does not need to be mapped to a physical qubit layout, as is required in QC architectures possessing solely next-neighbor connectivity \cite{corcoles_challenges_2020, farhi_quantum_2017, venturelli_compiling_2018}.

In order to encode both the spin-1/2 degree of freedom and the fermion occupancy of each spin on a site $i\in\mathbb{Z}$, we introduce an auxiliary pseudospin operator $I_i$, using the approaches in Refs. \cite{ruegg_z_2010, khodaeva_quantum_2024} which we briefly summarize here. The fermionic operator for each spin can now be written as \cite{ruegg_z_2010},
\begin{equation}
    \hat c_{i\sigma}^{(\dagger)} = 2 \hat I^x_i \hat f_{i\sigma}^{(\dagger)}
\end{equation}
with the pseudospins $I_i$ satisfying the standard SU(2) Lie algebra $[I_i, I_j] = i\epsilon_{ijk} I_k$, with $\epsilon_{ijk}$ being the fully antisymmetric Levi-Civita symbol. One route to implementing this on a quantum computer is to apply the Jordan-Wigner transformation directly to this encoding, thus requiring $2N$ qubits to represent $N$ sites \cite{ruegg_z_2010}. Another route is to transform the FHM Hamiltonian into the form of a lattice gauge theory, which we discuss next. This route is incentivized by using conserved additional local and global symmetries for error mitigation and post selection, described in the ensuing Sections \ref{sec:symmetry}, \ref{sec:post-selection-error-mitigation}. Note that the introduction of auxiliary pseudospins enlarges the Hilbert space beyond that of the original FHM \cite{khodaeva_quantum_2024}, and hence accompanying constraints must be applied to restrict the simulation to the physically relevant subspace of the enlarged Hilbert space.

\subsection{\label{sec:z2_lgt}\zz LGT Transformation}
To frame the FHM as a lattice gauge theory (LGT), we introduce a second auxiliary spin-1/2 operator $\hat \tau^z_{ij}$ on the bonds between adjacent spins $i$ and $j$  to frame the FHM as a lattice gauge theory (LGT):
\begin{equation}
    \hat \tau^z_{ij} = \hat I^x_i \hat I^x_j
\end{equation}
Using such a transformation, and assuming translational symmetry $J_{ij, \sigma}= J$, the \zz LGT Hamiltonian reads,
\begin{eqnarray}
H_{\rm LGT} &=& -4J \sum_{j, \sigma} \left(\hat \tau^z_{j, j+1} \hat f^\dagger_{j\sigma} \hat f_{j+1, \sigma} + {\rm H.c. } \right) + \nonumber \\
& & \ \ + \frac{U}{2} \sum_j \hat \tau^x_{j-1, j} \hat \tau^x_{j, j+1}
\end{eqnarray}
The major change that engenders this to be a \zz LGT is the mediation of the hopping between neighboring sites by a dynamical spin operator $\hat \tau$ with eigenvalues $\pm 1/2$, instead of a number $J_{ij}$. 


\subsection{\label{sec:jw}Mapping fermions to spins: Jordan-Wigner Transformation}
The final step in our encoding before trotterization for digital quantum simulation is the conversion of fermions to spin-1/2 operators while preserving the canonical fermionic anticommutation relations, which is effected by a Jordan-Wigner transformation  \cite{jordan_uber_1928, giamarchi2003quantum}, :
\begin{eqnarray}
    \hat f_{i\uparrow}^\dagger &=&\prod_{\ell<i} (-)^{ \hat n_{\ell\uparrow}}\  \hat S_{i\uparrow}^+ \nonumber\\
    \hat f_{i\downarrow}^\dagger &=&  (-1)^{\sum_{\ell=1}^{N} \hat n_{\ell\uparrow}}\ \prod_{m<i} (-)^{\hat n_{m\downarrow}} \  \hat S_{i\downarrow}^+
\end{eqnarray}
The number operators on each site $\hat n_{i\sigma}=\hat c_{i\sigma}^\dagger \hat c_{i\sigma}$ have eigenvalues $\{0,1\}$ whereas the spin operators have eigenvalues $\pm 1/2$. They are related by $\hat S_{i\sigma}^z = \hat c_{i\sigma}^\dagger \hat c_{i\sigma} - 1/2$. Note that in our encoding, the occupied fermionic state is represented by $|0\rangle $ and the unoccupied state is represented by $|1\rangle$ in the computational basis of the qubits, and these have eigenvalues $+1$ and $-1$ respectively when measured in the $Z$ basis. Thus, the explicit relationship between the quantum computer expectation values reads,  
\begin{equation}
    \langle Z_i \rangle = 2 \langle \hat n_{i\sigma} \rangle- 1 = 2\langle\hat S^z_{i\sigma}\rangle 
\end{equation}
\subsection{\label{sec:trotter}Trotterization of LGT Hamiltonian}
With the transformations above, the LGT Hamiltonian can be recast exclusively in terms of spin-1/2 operators with nearest-neighbor interactions between the three sets of spins, 
the two site spins $S_{i\uparrow}, \hat S_{i\downarrow}$ and the bond spin $\hat \tau_{i, i+1}$ linking each spin for adjacent sites.
\begin{eqnarray}
    H_{\rm LGT} = &-& 4J \sum_{i, \sigma} \left( \hat \tau^z_{i, i+1} \hat S^+_{i\sigma} \hat S^-_{i+1,\sigma} + {\rm H.c.} \right) \nonumber \\
    &+& \frac{U}{2} \sum_i \hat \tau^x_{i-1, i}\, \hat \tau^x_{i, i+1} 
    \label{eq:LGTspin}
\end{eqnarray}
Using a Trotter-Suzuki decomposition, the unitary evolution operator at time $t_n = n\Delta t $ can be written as $U = \exp(-iH_{\rm LGT} t) = \prod_{k=1}^n \exp({-iH_{\rm LGT} \Delta t})$, which is exact since $H_{\rm LGT}$ has no explicit time dependence. Each Trotter step can now be approximated as \cite{lloyd_universal_1996},
\begin{equation}
    e^{-i H_{\rm LGT} \Delta t} = e^{-i H_J \Delta t} e^{-iH_U\Delta t} + O((\Delta t) ^2)
\end{equation}
where we have identified the two terms with coefficients $J$ and $U$ from Eq. \ref{eq:LGTspin}, and the error term comes from the fact that these two terms are noncommuting. Based on Ref. \cite{khodaeva_quantum_2024}, this can be further decomposed as,
\begin{eqnarray}
    e^{-i H_J \Delta t} &  \approx &\prod_{j_{\rm even}\uparrow}\! \hat C_{j\uparrow} \! \prod_{j_{\rm odd}\uparrow} \! \hat C_{j\uparrow} \prod_{j_{\rm even}\downarrow} \hat C_{j\downarrow}  \prod_{j_{\rm odd}\downarrow} \hat C_{j\downarrow} \label{eq:HJ}
 \\
    e^{-iH_U\Delta t} & \approx & \prod_{j\uparrow}\hat B_{j\uparrow} \times \prod_{j} \hat B_{j} \label{eq:HU}
\end{eqnarray}
where $\hat C_{j\sigma} = \exp(4iJ \Delta t (\hat \tau^z_{j, j+1} \hat S^+_{i\sigma} \hat S^-_{i+1,\sigma}+ \rm H.c.))$ represents the hopping term and $D_j = \hat \tau^x_{j-1, j}\, \hat \tau^x_{j, j+1} $ represents the on-site interaction term. The ordering in the above equation is chosen so that the $C_{j_{\rm even}\sigma}$ operators for each spin $\sigma$ can all be implemented in parallel for $j$ being an even integer, followed by the same for $j$ being odd, as this operator involves three qubits: two adjacent site spins $\hat S_{j\uparrow}$, $\hat S_{j\downarrow}$ and the bond spins $\hat \tau_{j,j+1}$. The other, on-site interaction term $H_U$ is completely encoded in the bond spins, and hence $\hat B_{j}$ is directly implementable as a two-qubit quantum circuit. These implementation details are laid out in Fig. \ref{fig:explicit_circuits}  and elucidated below. Note that the Hubbard term that describes on-site interaction in the original Hamiltonian (Eq. \eqref{eq:FH}) has support only on nearest-neighbor bond qubits. 

\section{\label{sec:parameterized} Parameterized single trotter step ansatz}

\begin{figure}
    \centering
    \includegraphics[width=0.50\textwidth]{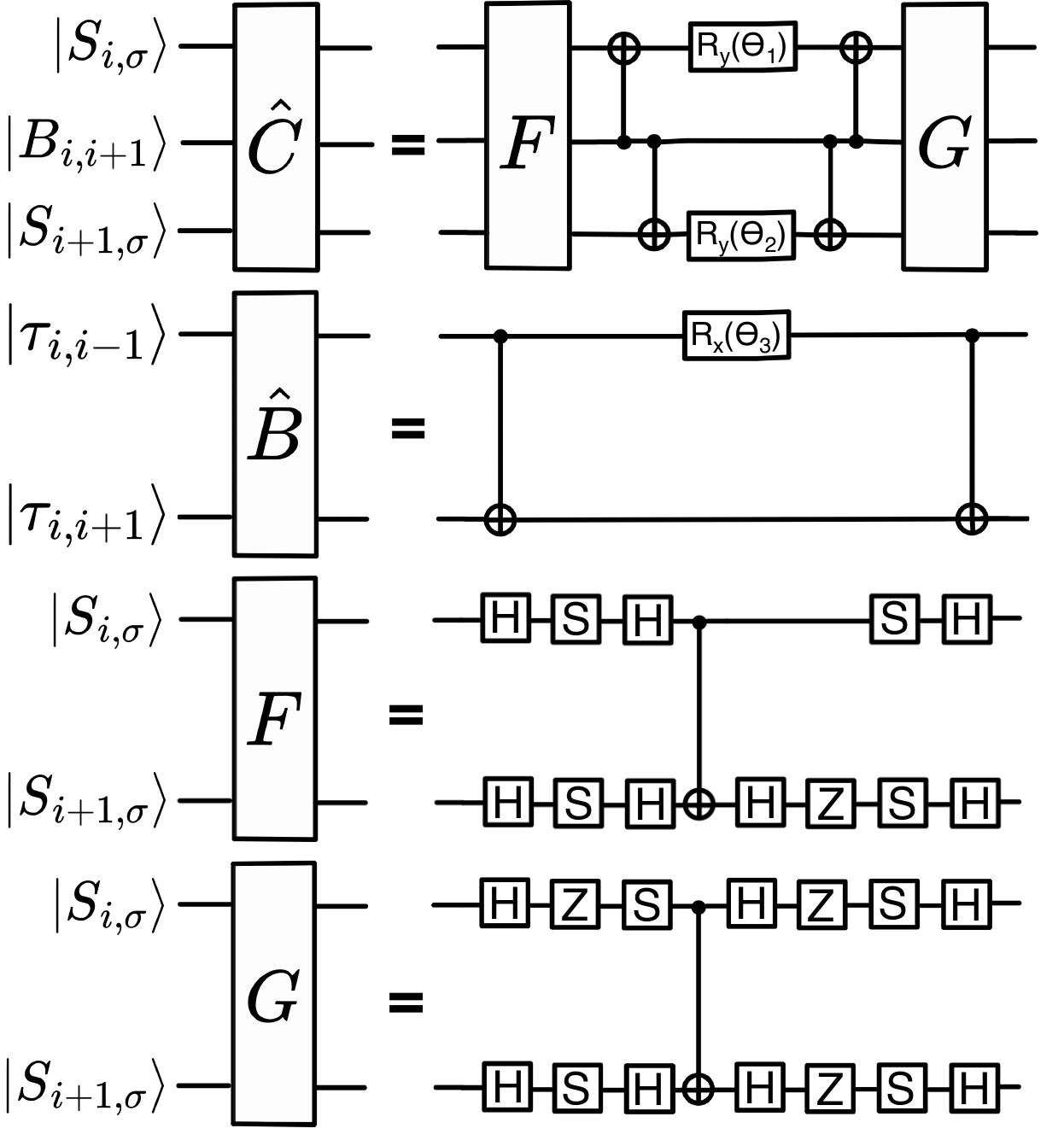}
    \caption{Direct circuit decomposition of $\hat{C}$ and $\hat{B}$ for a single Trotter step of $H_{LGT}$ from \cite{khodaeva_quantum_2024}. $\theta_{1} = 2J\Delta t$, $\theta_{2} = -2J\Delta t$, and $\theta_{3} = U\Delta t$}
    \label{fig:explicit_circuits}
\end{figure}

A direct implementation of the LGT-encoded FHM Hamiltonian's Trotter step was provided in Ref. \cite{khodaeva_quantum_2024}, which we use as our starting point before moving to the parameterized optimized ansatz in Sec. \ref{sec:parameterized}. Since the previous implementation was based on the IBM Quantum architecture using superconducting qubits, the universal gate set chosen includes a two-qubit entangling gate -- the CNOT gate -- and single qubit gates such as the Hadamard, Pauli $Z$, and the $S$-phase gate, along with arbitrary single qubit rotation gates $R_y(\theta) = \exp(-iY\theta/2)$ and $R_x(\theta) = \exp(-iX\theta/2)$. Using standard decompositions of two-qubit gates \cite{vatan_optimal_2004, iten_quantum_2016}, one arrives at the circuits shown in Fig. \ref{fig:explicit_circuits} for the $\hat C$ and $\hat B$ operators introduced in Eqs. \eqref{eq:HJ} and \eqref{eq:HU}. Although this has been thought to be an optimal decomposition \cite{vatan_optimal_2004}, our analysis in the next section establishes that the circuits can be substantially condensed, both in terms of single-qubit gates and more importantly in terms of two-qubit entangling gates, which holds significance in the noisy intermediate-scale quantum (NISQ) era.

In this section, we outline and implement a process to deconstruct a Trotter step into subroutines that can be optimized classically in a modular fashion, and subsequently recompose the Trotter step to generate an equivalent, parameterized variation of it.

\subsection{\label{sec:symmetry} Mapping FHM Hamiltonian's symmetries to a modular circuit ansatz}


Keeping the timestep $\Delta t$ during trotterized time-evolution of a system is recommended on a digital quantum computer. This means that each permutation of $\hat{C}$ only acts on one three-qubit subspace at a time, allowing for straightforward classical simulation of $\hat{C}$ and efficient reconstruction of each overall trotter step. Additionally, $\hat{C}$ occurs $2N$ times per Trotter step (where $N$ is the number of sites), compounding any improvement to $\hat{C}$ across an entire trotter step. This is especially true in the context of single and two-qubit gates: Two-qubit gate errors can exhibit one to three orders of magnitude higher infidelity than single qubit gates, even amongst recent state-of-the-art reports \cite{dasilva2024} \cite{Oliver_Fluxonium}, resulting in two-qubit gate errors dominating single-qubit errors at high gate counts. Therefore, any two-qubit gate reduction in $\hat{C}$ can increase the number of Trotter steps that can be reliably executed on a NISQ device. Section \ref{sec:von_Neumann} introduces a more systematic way to more aggressively decrease the two-qubit gatecount per Trotter step.

Imposing cyclic boundary conditions on the Fermi-Hubbard Model requires $\hat{C}$ to act on physically distant qubits on a chip. This is achievable on IonQ's trapped-ion systems without extra SWAP gates due to all-to-all qubit connectivity. 

Finally, quantum circuits submitted to cloud accessible QPUs undergo transpilation. Therefore, it is beneficial to simplify a circuit into a gateset that is easily mappable to the QPU in question. Here the concept of a native gateset is important, which we discuss in the Appendix, Sec. \ref{sec:ionq_QC}.

\section{Algorithm-Hardware Codesign of the Circuit}

In this section, we detail our approach that reduces the two-qubit gate count of the Trotterized \zz LGT Fermi–Hubbard model by up to 36\%.

\subsection{\label{sec:level2} Determining simplifiable subroutines}

To optimize $\hat{C}$, we create a parametrizable circuit (in number of gates and overall depth) $\hat{C}_{A}$, known as an ansatz, that satisfies $\hat{C}_{A} \approx \hat{C}$. Our ansatz is shown in Fig. \ref{fig:optimization}, and is comprised of single qubit $R_{X}(\theta)$ and $R_Y{\theta}$ gates, and a two qubit $XX(\theta)$ gate.

At a given circuit depth $l$, a gradient-based optimizer can find gate angles that satisfy $\hat{C}_{A} \approx \hat{C}$.

In each iteration of gradient-based optimization (GBO), gate angles are generated by an optimizer. Then, using the matrix representation of $\hat{C}$ and $\hat{C}_{A}$, the similarity between the two circuits is given by $\mathcal{F}$, where
\begin{equation}
    \mathcal{F} = \frac{\left( \sum_{ij} C_{ij}^* C_{A_{ij}}\right)^2}{\left( \sum_{ij}|C_{ij}|^2 \right)\left( \sum_{ij}|C_{A_{ij}}|^2  \right)} = \left( \sum_{ij} C_{ij}^* C_{A_{ij}}/ 2^{q}\right)^2 
\end{equation}
$\mathcal{F} = 1$ denotes $\hat{C} = \hat{C}_{A}$, and $\mathcal{F} = 0$ denotes $\hat{C} \perp \hat{C}_{A}$.

Converging to a solution is limited by local minima, barren plateaus, initial conditions, noise when computing the gradient, and the number of tunable parameters. A larger depth $l$ results in more parameters to optimize over and the introduction of further minima and plateaus in the optimization problem. Therefore, an optimizer that is consistent in performance and resilient to these challenges is critical to finding gate angles that satisfy $\hat{C} \approx \hat{C}_{A}$.

Accelerated gradient descent techniques are needed to find suitable gate angles that satisfy $\hat{C} \approx \hat{C_{A}}$. This is done by minimizing a cost function $f(x_{*}) = 1 - \mathcal{F}(x_{*})$, where $x_{*}$ is the gate angle vector. Formally, the goal is to compute a gate-angle vector $x_* \in \R^d$ such that
\begin{align}
    x_* \in X_* = \arg \min_{x \in \R^d} f(x). \label{eqn:opt_1_main}
\end{align} 

Newton's method~\cite{kelley1999iterative} offers a method of accelerating convergence by using the Hessian of $f$. Specifically, Newton's method uses preconditioning, which pre-multiplies the gradient with the inverse Hessian matrix at every iteration, resulting in local quadratic convergence rates~\cite{kelley1999iterative} for strongly convex objectives. Despite its faster convergence rate, there are several issues in Newton's method. (i) For empirical risk minimization, the per-iteration computational cost of Newton's Method is $\bigO(n d^2 + d^3)$. (ii) Secondly, the accelerated convergence of Newton's method is guaranteed only if $f$ is strongly convex, an unlikely condition in quantum circuit optimization. (iii) It involves computing a matrix inverse at every iteration, which is highly unstable against {\em process noise}, such as qubit decoherence and gate infidelity. While Newton's method offers superior convergence rates, these issues ultimately mean it is ill-suited for use in this setting.

\subsection{\label{sec:ipgshortdesc}
Iteratively Pre-conditioned Gradient Descent}

We present iteratively pre-conditioned gradient descent (IPG)~\cite{kushal20,chakrabarti2022iterative}, an accelerated gradient algorithm, to find suitable gate angles to surmount these challenges. IPG is a quasi-Newton method that iteratively estimates the inverse Hessian in a way that is more robust to process noise when compared to the explicit inverse Hessian and can remain stable even with a nonconvex cost. IPG achieves this by finding an approximate inverse Hessian with offset eigenvalues to help maintain stability, rather than explicitly taking the inverse Hessian like Newton's method. One notable difference with other quasi-Newton methods, such as BFGS~\cite{kelley1999iterative}, which also construct an approximation of inverse Hessian, is that the preconditioner in IPG is not required to be symmetric and positive definite, a condition likely to be violated when implemented with noisy quantum gates.

We have previously shown the utility of the IPG algorithm on quantum circuit optimization \cite{srinivasan_quantum_2023} whereby faster convergence was observed over other optimizers like gradient descent and Adam \cite{kingma_adam_2014}. Nevertheless, the exponential growth of the Hilbert space dimension with the number of qubits makes the problem of optimizing the entire quantum circuit simultaneously via IPG infeasible. Instead, we identified a 3-qubit unitary operator $\hat C$ which repeatedly appears at each Trotter step and optimize this subcircuit. By keeping the same timestep for each trotter step, the size of the circuit to be optimized stays fixed independent of the number of sites $N$ or qubit count $3N$. This leverages the model's translational symmetry discussed in Sec. \ref{sec:symmetry} and the all-to-all connectivity of the trapped-ion QPU which makes implementing periodic boundary conditions connecting $i=1$ to $i=N$ straightforward. Further details of IPG are discussed in the Appendix, Sec. \ref{sec:ipg}.

\subsection{\label{sec:opt_ansatz} Optimizing the Ansatz}
\begin{figure}
    \includegraphics[width=0.50\textwidth]{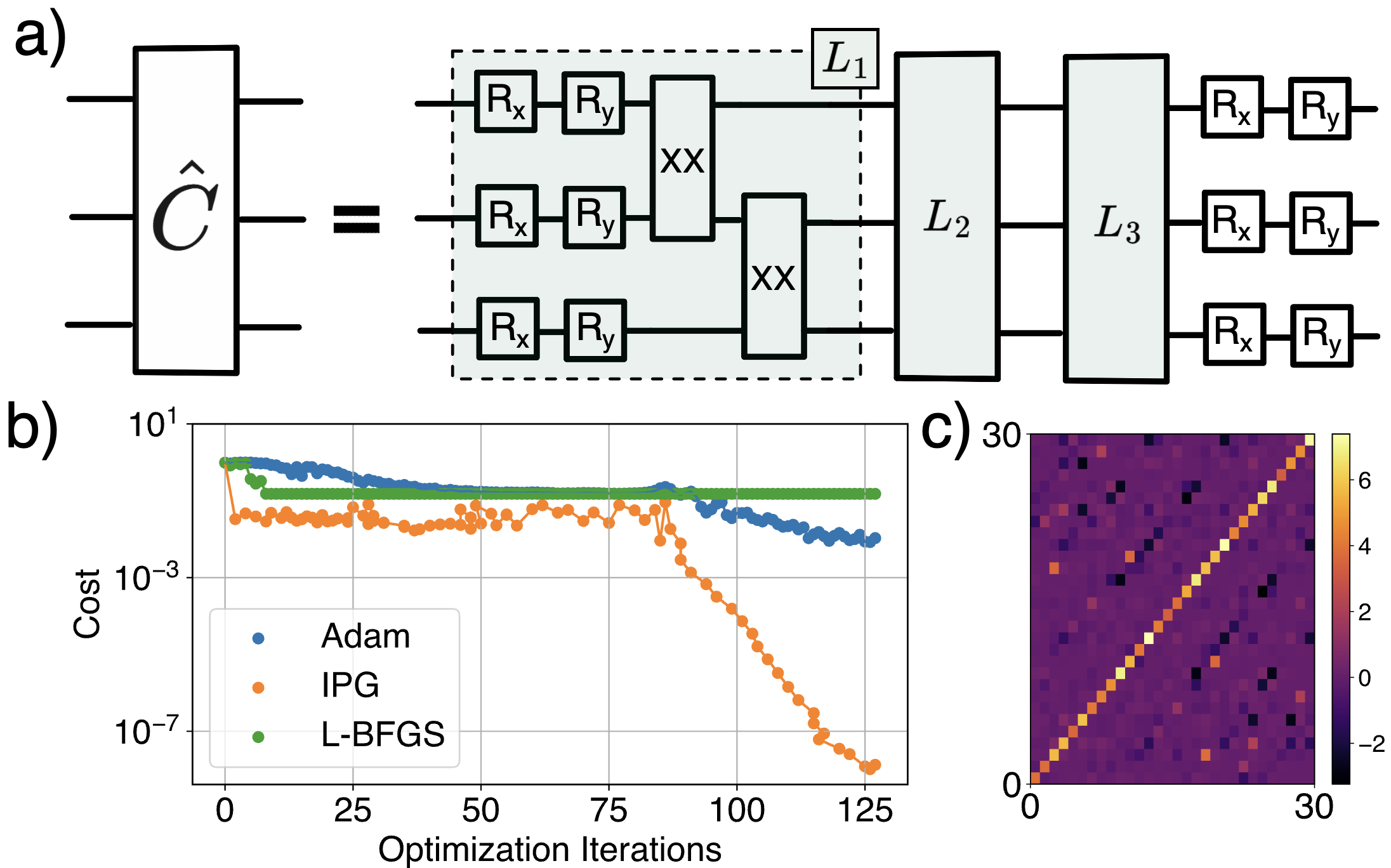}
    \caption{Gradient based optimization for circuit optimization. \textbf{a)} Modified ansatz $\hat C_{A}$ from \cite{kandala_hardware-efficient_2017} used for gradient based optimization to parametrize $\hat C$. Each layer is comprised of $R_{X}(\theta_{1})$, $R_{y}(\theta_{2})$, $XX(\theta_{3})$ on qubits $S_{i, \sigma}$, $B_{i, i+1}$, and $S_{i+1, \sigma}$ respectively. Three layers are used with a final $R_{X}(\theta)$ and $R_{Y}(\theta)$ rotation on each qubit. \textbf{b)} Result of GBO for 30 parameters using IPG, Adam, and L-BFGS. Best cost history over three trials and 128 iterations for each optimizer. IPG achieves the highest fidelity circuit, approaching fidelity greater than $\mathcal{F} = 1- 10^{-7}$. L-BFGS is trapped in a local minimum. Note that while there were variations depending on the target unitary $\hat C$ and the initial guess, these trends were found to be quite representative of the observed behavior during optimization. The optimized angle for the $XX(\theta) \approx \mathbf{I}$ between $|S_{i, \sigma} \rangle$ and $| S_{i+1, \sigma} \rangle$, allowing that gate to be removed. \textbf{c)} Final pre-conditioner matrix $K$ from IPG, which estimates the inverse Hessian. Matrix elements concentrated around the main diagonal indicate a well conditioned estimate of the the inverse Hessian and a stable optimization result.}
    \label{fig:optimization}
\end{figure}

We optimize the gate angles in $\hat{C}_{A}$ at a depth of $l = 3$ layers (Fig. \ref{fig:optimization}a) using IPG, and benchmark IPG performance to L-BFGS and Adam, two optimizers commonly used for quantum circuit optimization. Cost is defined as $1 - \mathcal{F}$, where lower cost indicates better agreement between $\hat{C}_{A}$ and $\hat{C}$.

Fig. \ref{fig:optimization} shows results for circuit optimization for each optimizer. IPG converges to the lowest cost in the fewest iterations, approximately four orders of magnitude lower than L-BFGS. To verify that the converged gate angles of each optimizer satisfy $\hat{C}_{A} \approx \hat{C}$, we apply 10000 random input states $\ket{\psi_{\rm in}}$ on both circuits and take the average of the state infidelity as
\begin{equation}
    {\rm Cost} = 1 - |\langle \psi_{\hat{C}_{A,i}} | \psi_{\hat{C}_{i}} \rangle |^2
\end{equation}
for each resulting output state $\ket{\psi_{C_{A,[i]}}} = \hat C_{A,[i]} \ket{\psi_{\rm in}}$. In doing so we find that only the gate angles generated by IPG are suitable enough to represent $\hat{C}$.

After constructing the overall Trotter step using $\hat{C}_{A}$ with angles converged to by IPG, we verify that the reconstructed trotter step is the same as the original, using the convergence method outlined above. The average difference between states generated by a single parameterized vs directly executed trotter step is of order $10^{-4}$, sufficiently low to consider both circuits approximately equal.

\subsection{\label{sec:von_Neumann} Decreasing two-qubit gatecount with von Neumann entropy}
In our above optimization, the number of layers ($l$) in $\hat{C}_{A}$ is a preselected hyperparameter.  Larger $l$ corresponds to a more expressive ansatz, at the cost of more gates. Ideally, we would like to minimize $l$ prior to applying IPG on $\hat{C}_{A}$.

\begin{figure}
    \centering
    \includegraphics[width=0.50\textwidth]{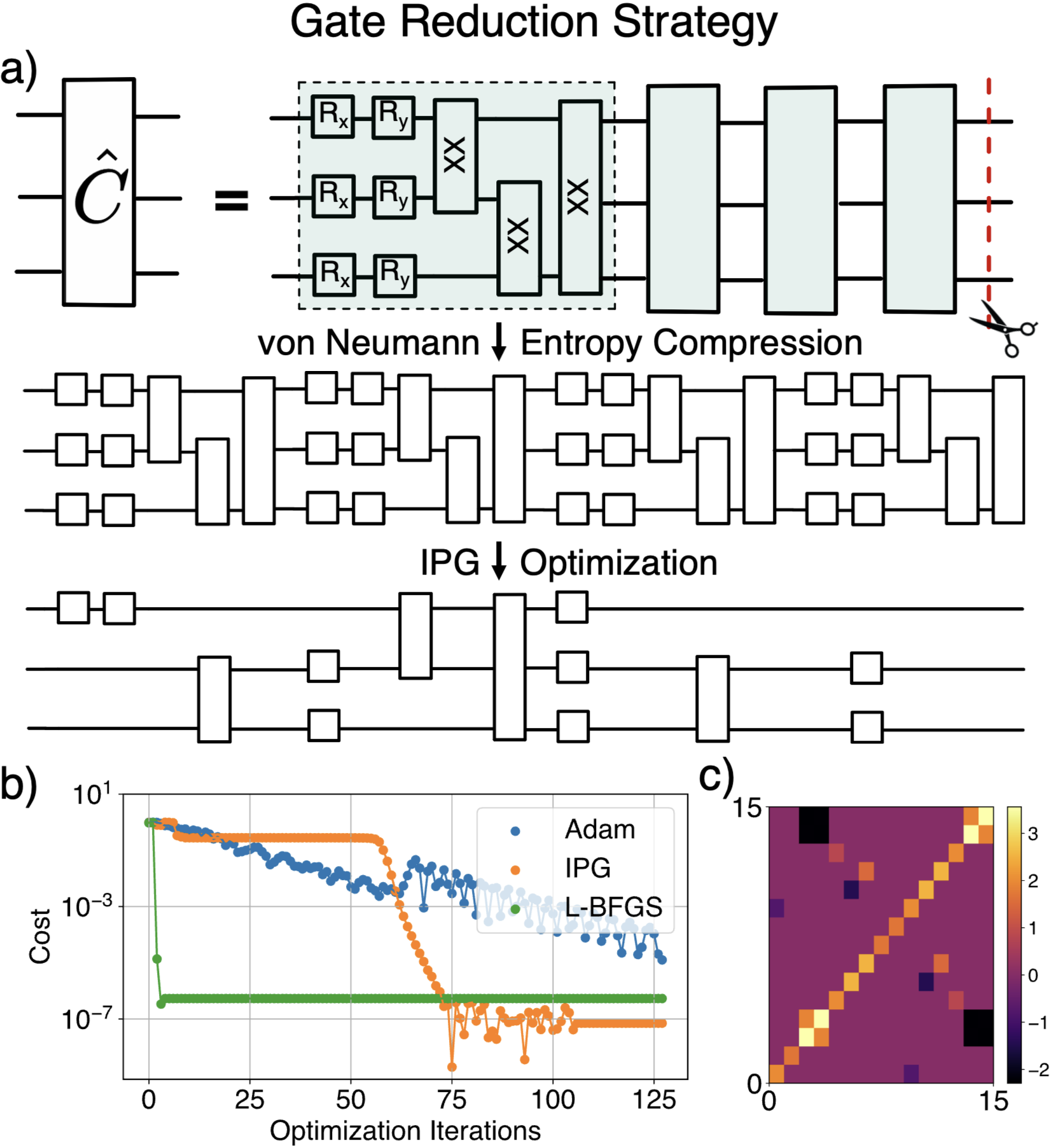}
    \caption{Combining von Neumann Entropy (VNE) optimization with gradient based optimization for further two-qubit gate reduction on each subcircuit. \textbf{a)} Ansatz for the VNE optimization method. The ansatz number of layers is first truncated through the VNE method. Following this, gradient based optimization (GBO) is applied to find parameters such that $\hat{C}_{A} \approx \hat{C}$.During GBO, a number of single and two-qubit gate rotations go to $\mathcal{I}$, allowing further compression of the ansatz. \textbf{b)} Result of GBO over 13 parameters using IPG, Adam, and L-BFGS. Best cost history over three trials and 128 iterations for each optimizer. IPG achieves the highest fidelity circuit, approaching fidelity of $\mathcal{F} = 1- 10^{-7}$, with L-BFGS converging faster but to a lower fidelity (higher cost). \textbf{c)} Final pre-conditioner matrix $K$ from IPG, which estimates the inverse Hessian, similar to Fig. \ref{fig:optimization}. 
    }
    \label{fig:classical_optim}
\end{figure}

To do so, we first modify the ansatz in Fig. \ref{fig:optimization} by adding an additional XX gate between $|S_{i, \sigma} \rangle$ and $|S_{i+1, \sigma} \rangle $. This creates the ansatz in Fig \ref{fig:classical_optim}, ensuring both full single-qubit rotations for $l \geq 2$ and possible entanglement between each qubit in each layer. Now, the number and placement of entangling gates are the limiting factors to determine whether $\hat{C}_{A}$ can express $\hat{C}$. By evaluating the von Neumann entropy (VNE) of different permutations of two-qubit gate placements in our ansatz, we can predict the smallest two-qubit gate depth necessary to satisfy $\hat{C}_{A} \approx \hat{C}$.

\subsubsection{\label{sec:level2} Procedure for predicting the smallest depth of an ansatz}

For a given state $|\psi_{i} \rangle$, the von Neuman entropy (VNE) is given by 
\begin{equation}
    S(\rho_{i}) = -Tr(\rho_{i} \log(\rho_{i}))
\end{equation}
where $\rho_{i}$ is the density matrix of the state $| \psi_{i} \rangle$. In a noiseless environment, VNE measures the degree of entanglement a qubit or set of qubits with those that are traced out.

For a given ansatz $\hat{C}_{A}$ with layers $l$, we first apply $\hat{C}_{A}$ on a random initial state $| \phi_{i} \rangle $ to yield resulting state $| \psi_{i} \rangle$. By doing this many times, we compute the average VNE of all permutations of subsystem of $| \psi_{i} \rangle$, and apply this same process with $\hat{C}$. If each corresponding VNE term for $\hat{C}_{A}$ at a given $l$ is greater than that for $\hat{C}$, we observe that $C_{A}$ is expressible enough to represent $\hat{C}$, and apply gradient-based optimization on that circuit.

\subsubsection{Outcome of VNE optimization}

With this method, $l \geq 3$ satisfied this expressibility criteria. Gradient-based optimization only converged to a solution at $l=4$, with the gates shown in Fig. \ref{fig:classical_optim} going to $\mathcal{I}$. The optimization sequence amounted to a thirteen 1Q gate reduction compared to $\hat{C}_{A}$, and more importantly, two fewer 2Q gates than $\hat{C}$. By also applying this process to $\hat{B}$, we were able to reduce $\hat{B}$ to an $XX(\theta)$ gate on $|\tau_{i-1,i} \rangle, |\tau_{i,i+1} \rangle$ and a single $R_{y}(\theta)$ on $|\tau_{i-1,j} \rangle$. 

Table \ref{tab:table2} summarizes the overall reduction in two-qubit gates per Trotter step, with an overall reduction from 14N to 9N. This decrease is substantial: Assuming 400 2Q gate executions as the Aria-1 algorithmic qubit limit \cite{Lubinski_Application-Oriented_Performance_Benchmarks_2021}, we are able to increase the number of faithful Trotter steps simulated on the Aria-1 QPU by $\approx50\%$, from 4 to 7 for $N=6$. 

\begin{table}[b]
    \caption{\label{tab:table2}%
    2-qubit gate (2Q) cost per trotter step for modelling $N$ sites of the FHM encoded as a \zz LGT. Results summarized for a) the initial direct trotter step decomposition b) by parametrizing $\hat{C}$ and applying gradient-based optimization (GBO) on $\hat{C}_{A}$, and c) through applying GBO with the non Neumann Entropy (VNE) method on $\hat{C}_{A}$ and a parametrized $\hat{B_{A}}$. Our final method is a 36\% reduction in 2Q cost per trotter step without comprising on exactness of the \zz LGT FHM encoding.}
    \begin{ruledtabular}
        \renewcommand{\arraystretch}{1.3}
        \begin{tabular}{lc}
            \textrm{Trotter Step Variant} & \textrm{2Q Cost} \\
            \colrule
            Direct & 14N \\
            GBO on $\hat C_A$ & 12N \\
            GBO and VNE on $\hat C_A$ and $\hat B_A$ & 9N \\
        \end{tabular}
    \end{ruledtabular}
\end{table}

\section{\label{sec:error-mitigation} Error Mitigation}

In this section we outline two main error mitigation strategies employed on our circuit executions: Compiler lever (active) error mitigation strategies, and post-circuit execution error mitigation by exploiting knowledge of extensive symmetries of the original Fermi-Hubbard Hamiltonian and of the \zz LGT encoding \textit{a priori}.

\subsection{\label{sec:debiasing}Debiasing and Sharpening}

Circuits submitted to IonQ QPU can opt to go through an error mitigation framework known as debiasing and sharpening \cite{maksymov2023enhancing}. The debiasing compiles the circuit-to-be-executed into various equivalent hardware implementations. Consequently, a non-trivial amount of the noise in the result corresponding to each specific hardware implementation will be uncorrelated.  In such a case,  one can further engage the sharpening post-processing filter.  Here, we briefly sketch the detailed explanation in \cite{maksymov2023enhancing}. 

Real-world qubits bear various kinds of noise. Many of these noise sources are non-homogeneous. For example, the noise on qubit $i$ probably differs from that on qubit $j$. The noise of $R_x$ gates differs from that of $R_z$ gates. For this reason, different implementations of the identical unitary $U$ typically give differently noisy results:
\begin{equation}
    \pi(\gamma(r_{u})) = u + \delta u_{\Gamma - inv} + \delta u_{\gamma}.
\end{equation}
Here $r_u$ represents a circuit compilation of unitary $u$. $\pi(u)$ corresponds to the implementation of circuit $r_u$. $\gamma$ represents the transformation between theoretically equivalent implementations.  We use $\delta u_{\Gamma-inv}$ to represent the implementation invariant noise, and $\delta {u_{\gamma}}$ to represent implementation dependent noise.

Experimentally, we measure the circuit implementation in the computational basis and obtain a histogram of the statistics of the observed qubit configuration. With this representation, we can write the effect of transformation between equivalent implementations as 
\begin{equation}
    \tilde{\mathbf{h}}_u = \mathbf{h}_u + \gamma h_{\Gamma - inv} + \delta h_{\gamma} \ \in \ \mathbb{R}^{+}(2^{n}).
\end{equation}
Here, we use $\tilde{\mathbf{h}}_u$ to represent the final observed sample statistics.  

If we estimate $\tilde{\mathbf{h}}_u$ over different non-correlated compilations, then the contribution of the implementation-dependent part will vanish. The effect of the implementation-independent noise is the depolarization noise corresponding to a uniform noise floor.  
For example, if the ideal measurement statistics is 
\begin{equation}
    \mathbf{h}_u = (0,...,1_{k},...,0_{2^{n}}),
\end{equation}
then the actual statistic over many implementations will become
\begin{equation}
\begin{split}
    \tilde{\mathbf{h}}_u &= \mathbf{h}_u + \gamma h_{\Gamma - inv} + \langle h_{\gamma} \rangle_{\Gamma} \\
    &= \mathbf{h}_u + \delta h_{\Gamma-inv} \\
    &= \left(\frac{\epsilon}{2^{n} - 1}, \ldots, (1 - \epsilon)k, \ldots, \frac{\epsilon}{2^{n} - 1}\right).
\end{split}
\end{equation}
This approach of breaking out the implementation of a unitary into different compilations is essentially the idea behind diversification. In this sense, the sharpening filter can be understood as a nonlinear filter applied to suppress the noise floor to an amount allowed by the statistics. In practice, we sample from different qubit-routing \cite{wagner2023improving} and apply gate (single-qubit and two-qubit) twirling \cite{cai2020mitigating} to generate different implementations. 

\subsection{\label{sec:post-selection-error-mitigation} Error mitigation based on global spin and local charge conservation}

The Fermi-Hubbard model intrinsically possesses global symmetries: the total number of fermions of each spin $\sum_j \hat n_{j\sigma}$ individually commutes with the Hamiltonian in Eq. \eqref{eq:FH}, $ \left[\sum_j \hat n_{j\uparrow}, \hat H\right] = \left[\sum_j \hat n_{j\downarrow}, \hat H\right] = 0 $. After the Jordan-Wigner transformation, this implies conservation of the total spin on each of the two sets of qubits on alternating sites (excluding the bond qubits), $\sum_j S^z_{j\sigma}$. Additionally, the lattice gauge theory encoding introduces ancillary bond qubits which must obey local charge conservation similar to quantum link models \cite{yang_observation_2020, khodaeva_quantum_2024, schweizer_fixed}: $\hat Q_j = (-)^{\hat S_{j\uparrow} + \hat S_{j\downarrow} }\, \hat \tau^x_{j-1,j}\, \hat \tau^x_{j,j+1}$. These conserved symmetries restrict the state of allowed spaces in the Hilbert space of $3N$ qubits for an $N$ site lattice based on the initial input state at $t=0$. In practice, due to gate infidelities and Trotter errors, the output of the quantum circuit does indeed have nonzero probabilities of going outside the conserved subspace. These states can be removed through post selection of the output histogram to restore the symmetries. In both the noisy statevector simulations and the QPU runs, we observed that the enforcing of global spin conservation in the up-spin and down-spin sector individually affected the observed quench dynamics significantly. As discussed in the next section and in Table 1, about 50\% of shots were discarded in most QPU runs for the 6-site lattice, which results in a large fraction of the results surviving post-selection as opposed to what is often observed on current quantum hardware where more than 90\% of the runs have to be discarded for deep circuits \cite{stanisic_observing_2022, arute2020}. 

\section{\label{sec:results} QPU and Noisy Statevector Simulator Results}

\begin{figure}
    \centering    \includegraphics[width=0.45\textwidth]{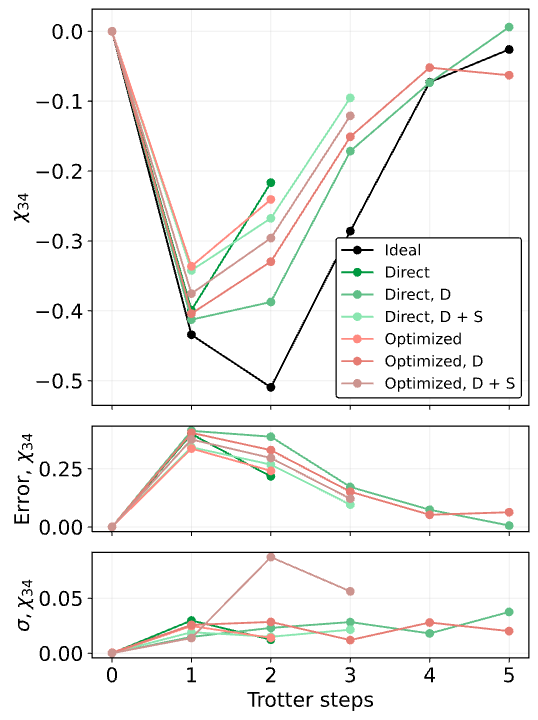}
    \caption{a) Circuit executions on the Aria-1 QPU without von Neuman entropy (VNE) two-qubit gate optimization (see Fig. \ref{fig:combined_classical} for results with VNE optimization). Magnetization correlator $\chi_{34}$ at five Trotter steps for $N=6$ sites of the Trotterized $H_{LGT}$ built from $\hat{C}$ (in various shades of green, with label "Direct") or $\hat{C}_{A}$ (in various shades of red, with label "Optimized"). Executed using 18 qubits. $J\Delta t = 0.3$, $U\Delta t = 0.6$. 
    Each circuit is subject the global spin conservation as a post selection error mitigation strategy. Each circuit goes through no compiler-level error mitigation, debiasing (D), or debiasing and sharpening (D + S). b) Error denotes the absolute value difference between the ideal result and a circuit execution at each Trotter step. c) $\sigma_{34}$ denotes the standard deviation in $\chi_{34}$ at each Trotter step across three circuit executions of each variant.}
    \label{fig:qpures}
\end{figure}

\begin{table}[b]
    \caption{\label{tab:table1}%
    Average root mean square error (RMSE) of $\chi_{i,i+1}$ between noiseless and QPU simulation for five Trotter steps of each circuit type run three times. Trotter step executions built from both the direct and optimized ansatz goes through either no compiler-level error mitigation strategy, debiasing (D), or debiasing and sharpening (D+S). Right column denotes the average number of shots discarded from post-selection error mitigation by removing states that do not satisfy total spin conservation.}
    \begin{ruledtabular}
    \renewcommand{\arraystretch}{1.3}
            \begin{tabular}{@{}lcc@{}}
            \textrm{Ansatz type}&
            \textrm{RMSE}&
            \multicolumn{1}{c}{\% Shots discarded}\\
            \colrule
                Optimized: D + S & 0.138 & 47.01\\
                Direct: D + S & 0.161 & 52.35\\
                Optimized: D & 0.167 & 52.66\\
                Direct: D & 0.147 & 53.87\\
                Optimized: None & 0.165 & 43.18\\
                Direct: None & 0.170 & 43.74\\
            \end{tabular}
        \end{ruledtabular}
\end{table}

This section presents our results from running the six-site Fermi-Hubbard model encoded using 18 qubits.

\subsection{Simulation parameters}

We Trotterize the LGT Hamiltonian with equal time steps. Site qubits are initialized in a separable state such that each site in the original FHM has one fermion, with the left half of the lattice being populated with $\uparrow$ spins and the right half with $\downarrow$ spins, creating a domain wall in the middle of the lattice. The $N$ bond qubits are initialized as  $\frac{1}{\sqrt{2}} (|+-\rangle ^{\otimes (N/2)} + |-+\rangle ^{\otimes (N/2)})$ 
for $N$ original fermion sites ($3N$ qubits). Due to cyclic boundary conditions, a domain wall also exists between site $i=1$ and site $i=N$. To benchmark circuit performance, we calculate quench dynamics using a two-point magnetization correlator $\chi_{ij}$ on opposite sides of the boundary between the up-spin domain and the down-spin domain (``domain wall") in each case,
\begin{equation}
    \chi_{i,i+1} = \langle \hat{S}_{i} \hat{S}_{i+1} \rangle - \langle \hat{S}_{i} \rangle \langle \hat{S}_{i+1} \rangle
\end{equation}
where $i=3$ for $N=6$, and $\hat{S}_i = \hat{n}_{i, \uparrow} - \hat{n}_{i, \downarrow}$

Circuits executed on the IonQ Aria-1 QPU are run in three ways: First without the error mitigation, then with debiasing, and finally with both debiasing and sharpening. All circuits are subject to post-selection spin conservation, such that only states in the histogram that possess the correct number of up and down spins are retained and the histogram is renormalized.

\subsection{\label{qpu-res}QPU Results}

For circuits executed on a QPU, subcircuits are optimized without the additional VNE optimization outlined in Sec. \ref{sec:von_Neumann}. Figs. \ref{fig:qpures} summarizes the evolution of the $\chi_{i,i+1}$ at the domain wall for five Trotter steps with $\Delta t = 0.3$, for $N = 6$ sites on the IonQ Aria-1 QPU.
Debiasing consistently yields a lower variance over multiple circuit executions of the same circuit for each Trotter step, indicating consistent performance across multiple identical circuit preparations. Sharpening improvess circuit execution consistency, with more pronounced improvement for the direct Trotter circuit compared to our optimized native gateset ansatz reconstruction. Overall, we see comparable variance and increase in variance as Trotter step increases for both the optimized and direct ansatz, with debiasing consistently decreasing this variance.

Table \ref{tab:table1} summarizes the accuracy of each circuit's $\chi_{i,i+1}$ evolution and the number of shots discarded as part of spin conservation post-selection. With the exception of evolution with debiasing, in which the directly implemented Trotter steps have a lower root-mean-square-error (RMSE) than the anstaz by 13\%, the remaining two cases have an average of 9.8\% lower RMSE from the optimized ansatz. Debiasing consistently improved circuit performance, with a more profound decrease in RMSE when applied to the direct circuits.

Prior studies \cite{khodaeva_quantum_2024} have evaluated the domain wall magnetization correlator on a noisy Qiskit simulator. These studies have used simulators that assume single and two-qubit depolarization as the sole error channel, with the magnitude of two-qubit depolarization error of order $10^{1}$ greater than single qubit error. In this study, all executed circuits remained within the average T1 time of each qubit used, but beyond T2 time, and at a two-qubit depth beyond the algorithmic qubit limit of the Aria-1 QPU \cite{Lubinski_Application-Oriented_Performance_Benchmarks_2021}. This, combined with the non-negligible impact of other noise channels, SPAM, and readout error, results in circuits performing poorer on a QPU than anticipated by currently available noisy simulators. In spite of this, the overall dynamics of the magnetization correlator are preserved, which is all the more remarkable in the face of $10^{-1}$ to $10^{-4}$ estimated overall circuit fidelity for our circuit executions, given an average 2Q gate fidelity of $\approx 98\%$ during our circuit executions.

Additionally, a prior work \cite{khodaeva_quantum_2024} has predicted that the number of shots per circuit run scales exponentially with $N$ for efficient post-selection. The result of this work is significantly more optimistic than this claim: Using a one-way analysis of variance (ANOVA) test at a significance level of $\alpha = 0.05$, we find no statistically significant difference in the percentage of shots discarded due to conserved symmetry-based error mitigation. This makes the number of shots discarded independent of whether debiasing and sharpening are used, despite debiasing and sharpening requiring an order-of-magnitude greater number of shots.

\subsection{\label{classical-res}Noisy statevector results}

\begin{figure}
    \centering
    \includegraphics[width=0.45\textwidth, keepaspectratio]{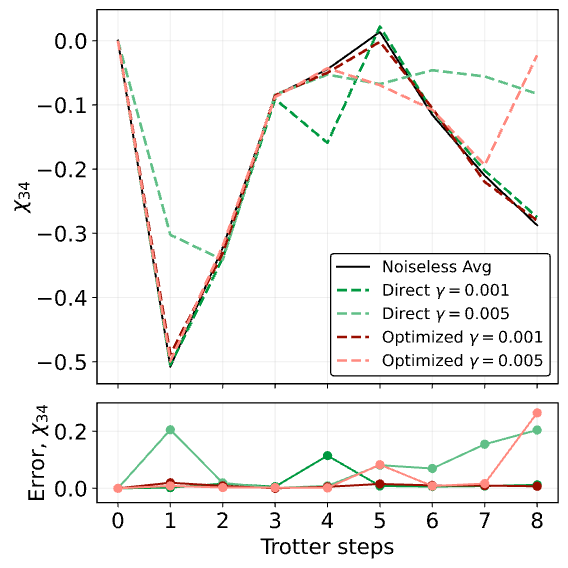}
        \caption{a) Circuit executions on a classical statevector simulator with von Neuman entropy (VNE) 2-qubit gate optimization. $\chi_{34}$ for $N = 6$ sites. Statevector simulator subject to a 2-qubit depolarizing noise channel at different weights $\gamma$. Magnetization correlator $\chi_{ij}$ at eight trotter steps of the trotterized $H_{LGT}$ built from $\hat{C}$ (green) or $\hat{C}_{A}$ (red). $J\Delta t = 0.4$, $U \Delta t = 0.8$. 1000 shots for every circuit execution. b) Error in $\chi_{ij}$ at each Trotter step is difference between each circuit and the noiseless result.}
    \label{fig:combined_classical}
\end{figure}

Subcircuits for our noisy statevector results were optimized using the VNE method outlined in Sec. \ref{sec:von_Neumann}. The statevector simulator is subject to a two-qubit depolarizing noise channel defined as
\begin{equation}
    E(\rho) = (1 - \gamma)\rho + \gamma Tr \left(\rho \frac{I}{2^{n}}\right)
\end{equation}
where $\rho$ is a density matrix, $\gamma$ is the depolarizing error parameter (where higher $\gamma$ denotes larger probability of error on $\rho$), and $n = 2$ qubits. 

Fig. \ref{fig:combined_classical} summarizes our results. We find an excellent representation of the dynamics of $\chi_{34}$ across seven trotter steps by the Trotter circuit constructed by the optimized ansatz for both $\gamma = 0.001$ and $0.005$. This is in sharp contrast to the directly constructed Trotter circuit, which loses a faithful representation of $\chi_{34}$ past just two steps at $\gamma = 0.005$ and four steps at $\gamma = 0.001$.

\section{\label{sec:conclusions} Conclusion}
In this work, we have presented an algorithm-hardware co-design methodology for executing deep, Trotterized digital quantum simulations of strongly-correlated systems on digital quantum computers. Our methodology is the integration of four core components: First, symmetry-constrained three-qubit sub-circuits that are parametrized and classically optimized via iteratively pre-condition gradient descent (IPG); second, expressibility guided two-qubit gate-count reduction using the von Neumann Entropy (VNE) of each sub-circuit; third, general purpose compiler-level error mitigation via debiasing and sharpening; and fourth, post-selection based error mitigation specific to the symmetry of the model of interest.

The specific instance we implemented is a six-site \zz LGT mapping of the Fermi-Hubbard Model (FHM) on a trapped-ion quantum computer, whereby periodic boundary conditions were readily realized by harnessing all-to-all qubit conenctivity. Our metholody yielded substantial circuit compilation gains: IPG optimization on a three-qubit subcircuits alone reduced the native two-qubit gate count per Trotter step by 14\%. This, combined with compiler-level debiasing and symmetry-based post-selection from global spin conservation error mitigation allowed for a faithful observation of domain-wall quench dynamics even at total circuit fidelities between $10^{-1}$ to $10^{-4}$. VNE compression yielded an additional 12\% two-qubit gate count reduction, bringing the total reduction to 36\%. This doubled the number of Trotter steps that could be simulated under a depolarizing noise channel on a clasical statevector simulator.


Our results have significant potential to be extended to more generalized versions of the Fermi-Hubbard model such as the SU(N) case, and those with extended interactions beyond on-site terms, whereby richer dynamics and phase transitions are anticipated \cite{hofstetter_high-temperature_2002, taie_observation_2022, pasqualetti_equation_2024}. We anticipate that the strategies showed in this work can be applied to other quantum many-body models beyond the Fermi-Hubbard Hamiltonian that host strongly correlated phases of matter \cite{preiss_strongly_2015, fauseweh_quantum_2024, smith_simulating_2019}, and to digital quantum simulation beyond condensed matter physics. 

\section*{Acknowledgements}
This work was supported by a QLab seed grant from the University of Maryland and IonQ, and an NSF CAREER award (2340835). We acknowledge discussions with Franz Klein and John Sawyer from the QLab.

\newpage

\bibliography{MyLibrary, manualrefs, refs_ipg} 

\section{Appendix}
\subsection{\label{sec:ionq_QC}Hardware constraints of a trapped-ion quantum processor}
\subsubsection{Trapped-Ion Native Gateset}
When a circuit representation of a quantum computation is submitted to an IonQ QPU via the default API, it goes through a compilation pipeline.  First, the circuit will be transpiled into a standard universal gate set. Then it is optimized using a quantum circuit optimizer (QCO) specifically designed for such a gate set. After the optimization, the circuit is converted to native gates and executed. In this sense, native gates are the unitary transformations implemented directly as tailored laser pulses acting on the ions, without further gate-based decomposition. 

IonQ’s native gates API allows direct access to two single-qubit gates and various two-qubit entangling gates depending on the specific system. The two single-qubit gates are referred to as $GPi(\phi)$ and $GPi2(\phi)$ :

\begin{equation}
GPI(\phi) = \begin{pmatrix}
0 & e^{-i\phi} \\
e^{i\phi} & 0
\end{pmatrix}
\end{equation}

\begin{equation}
GPI2(\phi) = \frac{1}{\sqrt{2}} \begin{pmatrix}
1 & -ie^{-i\phi} \\
-ie^{i\phi} & 1
\end{pmatrix}
\end{equation}

These are $\pi$ and $\pi/2$ rotations of the qubit along an axis that is specified by $\phi$ on the equatorial plane of the Bloch sphere \cite{wright2019benchmarking}. 

On the 32-qubit Aria systems, the supported native two-qubit gate is referred to as the arbitrary angle Molmer-Sorensen (MS) Gate \cite{wright2019benchmarking}. 
This is a simultaneous rotation of the target qubit 0 and target qubit 1 along the axis specified by $\phi_0$ and $\phi_1$ on the equatorial plane of the Bloch sphere, respectively.  
With full control over the provided degrees of freedom, these three gates make a universal gate set. 

Native gates are popular for a few reasons. In quantum characterization, verification, and validation (QCVV) studies, the native gates API allows bypassing optimizations. This is crucial, for example, to echo-type experiments that would otherwise be interrupted by optimization. 

The native gates also provide a pathway to improve the efficacy of circuit executions. For example, a $ZZ(\theta)$ type gate, which is by default implemented as two maximally entangling $MS(\pi/2)$ gates and one $R_z(\theta)$ rotation, can be implemented using just a single $MS(\theta)$ gate. This leads to a non-trivial improvement since entangling gates are an important source of noise. 
We note that for typical implementations of MS gates, the magnitude of noise is positively correlated with the gate angle. For example, a study using one of the now-decommissioned system \cite{nam2020ground} reported improvement from ~97.5\% to ~99.6\% two-qubit fidelity when comparing gate angles $\theta$'s  of $\pi/2$ and $\pi/100$, respectively. This makes native gates particularly preferable when a non-negligible small-angle entangling gate is required for the computation.  

\subsection{Iteratively Pre-Conditioned Gradient Descent} \label{sec:ipg}
This section introduces the Iteratively Preconditioned Gradient (IPG) descent algorithm, which computes a minimizer of the {\em cost function} $f:\R^d \to \R$. Formally, the goal is to compute a gate-angle vector $x_* \in \R^d$ such that
\begin{align}
    x_* \in X_* = \arg \min_{x \in \R^d} f(x). \label{eqn:opt_1}
\end{align}
When the cost function is non-convex, instead of searching for a globally optimal solution, a more meaningful and attainable goal is to find a {\em local minima} of the cost function $f$, defined as a point such that, for some $\epsilon > 0$ every point satisfying $|x^* - x|_2 < \epsilon$ must also satisfy $f(x) > f(x^*)$,
where $\nabla f(x) \in \R^d$ denotes the gradient of $f$ at $x \in \R^d$.

Built upon the prototypical gradient-descent (GD) algorithm~\cite{bertsekas1989parallel}, several accelerated gradient algorithms have been proposed for solving~\eqref{eqn:opt_1}~\cite{nesterov27method, polyak1964some, kingma2014adam, NEURIPS2020_d9d4f495}. Amongst them, some notable algorithms are Nesterov's accelerated gradient-descent (NAG)~\cite{nesterov27method}, the heavy-ball method (HBM)~\cite{polyak1964some}, and adaptive momentum (Adam)~\cite{kingma_adam_2014}. Although these methods improve GD's convergence rate, they converge at a {\em sublinear} rate~\cite{tong2019calibrating,su2014differential} for general nonconvex costs. 

By contrast, Newton's method~\cite{kelley1999iterative} offers another method of accelerating convergence by using the Hessian of $f$. Specifically, Newton's method uses preconditioning, which pre-multiplies the gradient with the inverse Hessian matrix at every iteration, resulting in local quadratic convergence rates~\cite{kelley1999iterative} for strongly convex objectives. Despite its faster convergence rate, there are several issues in Newton's method. (i) For empirical risk minimization, the per-iteration computational cost of Newton's Method is $\bigO(n d^2 + d^3)$. (ii) Secondly, the accelerated convergence of Newton's method is guaranteed only if $f$ is strongly convex, an unlikely condition in quantum circuit optimization. (iii) It involves computing an explicit matrix inverse at every iteration, which is highly unstable against {\em process noise}, such as qubit decoherence and gate infidelity. While Newton's method offers superior convergence rates, these issues ultimately mean it is ill-suited for use in this setting.

Iteratively Preconditioned Gradient Descent (IPG) is a quasi-Newton method that constructs an estimate of the inverse Hessian that is more robust to process noise when compared to the explicit inverse Hessian and can remain stable even with a nonconvex cost. The steps of the algorithm are shown in Algorithm \ref{alg:ipg}. Next, we briefly detail the Iteratively Pre-conditioned Gradient descent (IPG) algorithm. Like gradient descent and Newtons Method, IPG is an iterative optimization algorithm where in each step $t \in \{0, \, 1, \ldots\}$, an estimate $x_t \in \R^d$ of a minimum point Eq.~\eqref{eqn:opt_1} and a pre-conditioner matrix $K_t \in \R^{d \times d}$ are updated using steps presented below.

\begin{algorithm}[H]
    \caption{Iteratively Preconditioned Gradient Descent Algorithm}
\begin{algorithmic}[1] \label{alg:ipg}
    \Require Starting at $x_0 \in \mathbf{X}$, objective function $f$, initial preconditioner estimate $K_0$, stabilization parameter $\beta \geq 0$, initial parameter descent rate $\delta > 0$ 
    \For{$t = 0$ to $n$}
        \State $x_{t+1} \gets x_t - \delta_t K_t \nabla f_t$
        \State $\delta \gets 1/\left(\lambda_{\rm max}(\nabla^2 f_t) + \beta\right)$
        \State $K_{t+1} \gets K_t - \delta(\nabla^2 f_t K_t + \beta K_t - I)$
    \EndFor
\end{algorithmic}
\label{alg:ipg}
\end{algorithm}

The IPG algorithm is structured similarly to Newton's method in that both approaches rely on preconditioning the gradient to accelerate convergence. Still, they differ in the way the preconditioner is constructed. Specifically, rather than explicitly taking the inverse Hessian like Newton's method, IPG finds an approximate inverse with offset eigenvalues to help maintain stability. The notable difference with quasi-Newton methods, such as BFGS~\cite{kelley1999iterative}, which also construct an approximation of inverse Hessian, is that the preconditioner in IPG is not required to be symmetric and positive definite, a condition likely to be violated when implemented with noisy quantum gates. Next, we present the steps of Algorithm \ref{alg:ipg}.

{\bf Initialization:} Before starting the iterations, an initial estimate $x_0$ and a pre-conditioner matrix $K_0$ is chosen from $\R^d$ and $\R^{d \times d}$, respectively. Further, three sequences of non-negative scalar parameters $\{\alpha_t, \beta_t, \delta_t, t\geq 0\}$ are chosen, such that $\delta_t \leq 1$, $\beta_t > - \lambda_{\min} (H_t)$, and $\alpha_t < 1 / (\lambda_{\max} (H_t) + \beta_t)$. Here, $\lambda_{\min} (\cdot)$ and $\lambda_{\max} (\cdot)$ respectively denote the smallest and the largest eigenvalue of a square matrix.



{\bf Steps in each iteration $\bm{t}$}: For each iteration $t \geq 0$, we let $f_t = f(x_t)$, $g_t = \nabla f(x_t)$, and $H_t = \nabla^2 f(x_t)$ respectively denote the value of the cost function $f$, its gradient vector, and the Hessian matrix evaluated at the current estimate $x_t$.
Let $I$ denote the $(d \times d)$-dimensional identity matrix.
In each iteration $t$, the algorithm comprises two steps. 
In {\em Step 1}, the estimate $x_t$ is updated to $x_{t+1}$ such that
    \begin{align}
        x_{t+1} & = x_t - \delta_t K_t \nabla f_t. \label{eqn:x_update}
    \end{align}
In {\em Step 2}, the pre-conditioner matrix $K_t$ is updated to $K_{t+1}$ using second-order information $H_t$:
    \begin{align}
        K_{t+1} & = K_t - \alpha_t \left(\left(\nabla^2 f_t + \beta_t I\right)K_t - I\right). \label{eqn:K_update}
\end{align}

In deterministic settings, the convergence analysis of the IPG algorithm can be found in~\cite{chakrabarti_accelerating_2021}. In the presence of noise, $x_t$ in the IPG algorithm converges to a neighborhood of a stationary point $x_{st}$~\cite{chakrabarti_accelerating_2021-1}. Empirically, for solving classical optimization problems, IPG requires fewer iterations to reach the desired accuracy and obtains a smaller steady-state error than prominent gradient-based optimizers~\cite{chakrabarti2022control}. 

The aforementioned IPG algorithm has recently been utilized for optimizing gate-based quantum circuits~\cite{srinivasan_quantum_2023}. Specifically, IPG showed faster convergence to a higher fidelity for quantum state preparation and quantum algorithmic subroutines, such as approximating the full unitary matrix of a quantum Fourier transform and quantum state preparation. However, these results with IPG in~\cite{srinivasan_quantum_2023} were generalized for most gate-based QPUs. Particularly, the hardware-agnostic approach in~\cite{srinivasan_quantum_2023} was insufficient to benchmark the optimized circuits on the Aria-1 QPU. In this paper, we investigate the performance of IPG on optimizing hardware-aware (IonQ's trapped-ion) ansatz presented in Section~\ref{sec:parameterized}.

\end{document}